# Low Threshold Bound State in the Continuum Lasers in Hybrid Lattice Resonance Metasurfaces

Jhen-Hong Yang[1], Dmitrii N. Maksimov[2,3], Zhen-Ting Huang[4], Pavel S. Pankin[2,3], Ivan V. Timofeev[2,3], Kuo-Bing Hong[4], Heng Li[4], Jia-Wei Chen[4], Chu-Yuan Hsu[4], Yi-Yun Liu[5], Tien-Chang Lu[4], Tzy-Rong Lin[6], Chan-Shan Yang[7,8], Kuo-Ping Chen[9,*]

[1] *Institute of Photonic System, College of Photonics, National Chiao-Tung University, Tainan 71150, Taiwan*

[2] *Kirensky Institute of Physics, Federal Research Center KSC SB RAS, 660036, Krasnoyarsk, Russia*

[3] *Siberian Federal University, 660041, Krasnoyarsk, Russia*

[4] *Department of Photonics, College of Electrical and Computer Engineering, National Chiao-Tung University, Hsinchu 30010, Taiwan*

[5] *Institute of Lighting and Energy Photonics, College of Photonics, National Chiao-Tung University, Tainan 71150, Taiwan*

[6] *Department of Mechanical and Mechatronic Engineering, National Taiwan Ocean University, Keelung 20224, Taiwan*

[7] *Institute and Undergraduate Program of Electro-Optical Engineering, National Taiwan Normal University, Taipei 11677, Taiwan*

[8] *Micro/Nano Device Inspection and Research Center, National Taiwan Normal University, Taipei 106, Taiwan*

[9] *Institute of Imaging and Biomedical Photonics, College of Photonics, National Chiao-Tung University, Tainan 71150, Taiwan*

[*] *Corresponding Author: kpchen@nctu.edu.tw*

**ABSTRACT:** Bound states in the continuum (BICs) have attracted much attention in recent years due to the infinite quality factor ($Q$-factor) resonance and extremely localized field. In this study, BICs have been demonstrated by dielectric metasurfaces with hybrid surface lattice resonance (SLR) in the experiment. By breaking the symmetry of geometry, SLR can be easily switched between BICs and quasi-BICs. Comparing with literature, switching between BICs and quasi-BICs is usually accompanied by wavelength shift. Here, a design rule is proposed to prevent the wavelength shift when the $Q$-factor is changing. Also, such a design also makes subsequent identification of the laser threshold more credible. Due to the high $Q$-factor, low threshold laser is one of the intuitive applications of BICs. Utilize the high localized ability of BICs, low threshold BICs laser can be achieved by the dielectric metasurface immersed with Rhodamine 6G. Interestingly, due to the high $Q$-factor resonance of BICs, the laser signals and images can be observed in almost transparent samples. Not only the BICs laser is demonstrated in the experiment, but also the mechanism of BICs is deeply analyzed. This study can help readers better understand this novel feature of BICs, and provide the way for engineer BICs metasurfaces. The device can provide various applications, including laser, optical sensing, non-linear optics enhancement, and single-photon source.

In recent years, several works perform that simple photonic structures possess bound states above the threshold of continuum energy level[1-3]. Bound states in the continuum (BICs) are already demonstrated theoretically by von Neumann and Wigner in 1929[4]. These peculiar states possess infinite lifetime and the property of lossless systems. Therefore, BICs have attracted much attention due to its strong localization property and infinite quality factor ($Q$-factor). To the best of our knowledge, the first observation of the effect in the experiment is probably by Henry and co-authors in 1D grating[5], although they did not call it BICs those days. Strictly speaking they were working with a quasi-BICs (Q-BICs) as the true BICs in the system was destroyed by the finiteness of the structure[6]. Later on, a similar effect was observed in 2D photonic crystal[7] also with no reference to BICs. After publication of 2016 review by Chia Wei Hsu and co-authors[3] the researchers started to reference BICs much more often. Due to the fundamental mechanism of metasurfaces and photonic crystal are similar, many works are also presenting BICs by metallic[8-10] or dielectric[11,12] metasurfaces nowadays. Based on the symmetry protection, BICs can be switched to Q-BICs by breaking the symmetry of metasurfaces. However, breaking the symmetry arbitrarily leads to the obvious shift of resonance wavelength,[1,13] which is not conducive to dispersion material or further BICs applications.

The applications of BICs have been widely discussed in many fields, especially topology photonics[12,14,15]. By utilizing topological design to create BICs, different kinds of metasurfaces or photonic crystals can achieve abundant applications, *e.g.* photonic integrated circuits[16-19], sensor[20-22], nonlinear effect enhancement[23-26], low threshold laser[27,28], vortex laser[29-32]. Remarkably, the optical BICs have attracted attention not only as resonators for lasing but also as a source of thermal emission[33]. Among these applications, low threshold laser is intuitive for BICs applications due to its infinite $Q$-factor. In literatures, laser applications of BICs have been already presented by the dark resonance of two-dimensional periodic array nanostructures previously,[34,35] although they did not mention BICs at that time. However, due to the lossless of BICs, the lasing direction is not in the normal direction (not at the Γ point). Despite people can use symmetry breaking to generate radiation channel (radiation loss) in the normal direction,[36,37] the exact design rule and the mechanism behind the symmetry breaking are still not clear. Fortunately, there are several literatures of BICs lasers with metasurfaces are proposed.[27,31] Among these literatures, most of them put the gain medium inside the nanostructure or directive using gain medium to fabricate nanostructures. However, due to the decay of the gain medium, these metasurfaces with weakening gain medium is difficult to be reused. In addition, the BICs demonstrated by hybrid surface lattice resonance have not been proposed before.

In this article, $Si_3N_4$ metasurfaces with hybrid lattice resonance demonstrate the BICs by the complete dark state. To clearly show the design rule and the mechanism behind BICs, both simulations and experiments results reveal the mechanism of BICs by external excitation and internal excitation. Utilize internal excitation and the enhancement of electric dipole lattice resonances (EDLR), dielectric metasurface low threshold lasers can be achieved with Rhodamine 6G (R6G) and BICs at room temperature. Due to the design of EDLR, normal direction lasing (at $\Gamma$ point) can also be achieved in this study. In addition, because of the gain medium is outside the nanostructure and easy to be replaced, it makes the decayed laser device able to be reused. Moreover, the $Q$-factor of lasing mode can be tuned by changing the particle size, which means that the threshold and output efficiency can also be controlled by breaking symmetry. For further analysis of the laser threshold, in the process of changing the $Q$-factor, the shift of resonance wavelength can be prevented by the design rule in this article.

**DESIGN AND SIMULATION**

In the past several years, Fano resonance has been deeply studied, especially the high $Q$-factor resonance can be achieved by topological design[37-40]. When the Fano resonance occurs, destructive and constructive interference create the Fano shape curve in the spectrum. Once the destructive and constructive interference happen at the same wavelength, the clap of these resonances generates a flat shape in the spectrum but possesses infinite $Q$-factor, also called BICs[11,41,42]. To present the characteristic of BICs, the periodic structure is intuitive to be utilized, which is because topological photonics have been widely demonstrated by periodic structure[43-45]. Among many periodic structure designs, nanoparticle array has been widely applied to create additional nanostructure resonances, also called surface lattice resonance (SLR), which occurs at wavelengths close to Rayleigh anomalies (RAs)[46-49]. The SLR can be excited by a plane wave, which is illuminated by the normal direction of the surface lattice. Once the SLR is generated, constructive interference between each particle can create a strong resonance peak or dip in the spectrum from the far-fields. In order to create BICs, high $Q$-factor Fano resonance can be achieved by two surface lattice resonances (SLRs) which possess $\pi$ phase difference. Two anti-phase SLRs can be generated by the hybrid lattice system, which has been demonstrated theoretically in the literature[50]. In our study, hybrid SLR is generated by $Si_3N_4$ nano brick arrays, which is due to the characteristics of high refractive index and low material loss is necessary for creating BICs.

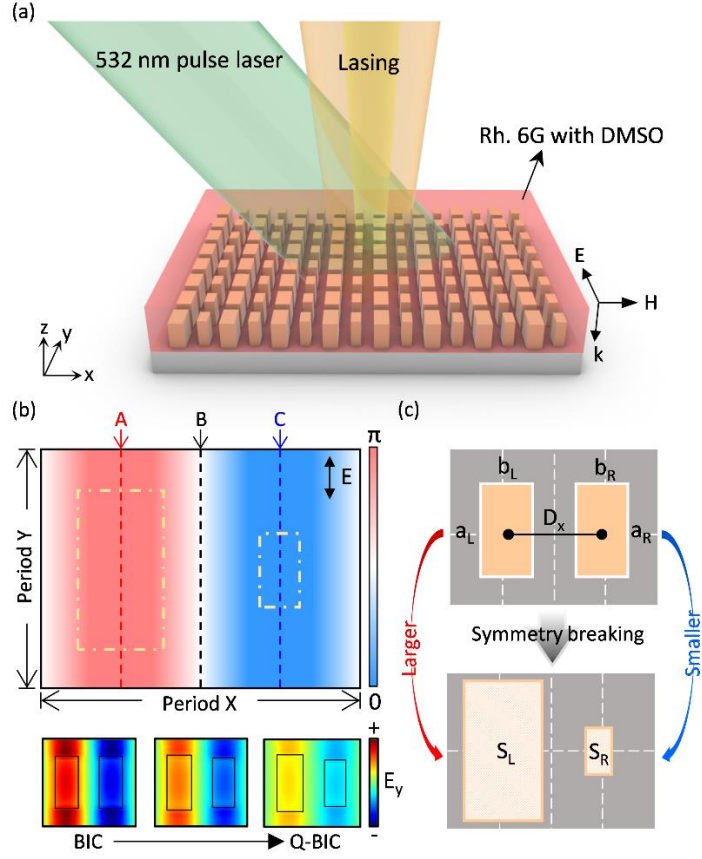

**Figure 1.** (a) Schematics of $Si_3N_4$ hybrid lattice metasurfaces with R6G. (b) The phase distribution color map of lattice resonance in *y*-polarization. The period of x-direction is 400 nm, and that of y-direction is 300 nm. (c) The design for breaking symmetry of $Si_3N_4$ hybrid surface lattice.

**Figure 1a** shows the schematic of $Si_3N_4$ hybrid lattice structure with photoluminescence (PL) material (R6G). In this article, electric dipole lattice resonance (EDLR) is utilized to generate BICs, which is because the electric field is localized outside the structure, thus the PL material would cover the $Si_3N_4$ metasurface for the lasing. Due to the electric filed polarization is in *y*-direction, the phase variation is just presented in *x*-direction at the wavelength of SLR, which can also be observed in **Figure 1b**. In order to explain the mechanism of BICs, the phase difference between positions A, B, C should be explained. If particles of the hybrid lattice are placed at position "A and B" or "B and C", the two independent SLRs can be created and without any interference. In contrast, if particles are placed at positions A and C, the two SLRs are created with strong interference in the far-field. Note, the description above is based on that there is not near-fields coupling effect between these two particles. Utilize the phase difference between different positions, BICs (completely dark resonance) can be created when there are two identical particle at position A and position C. Due to the infinite $Q$-factor of BICs, the electric field is localized at the plane of nano brick arrays. In the bottom of **Figure 1b**, the electric field intensity is decreased gradually by breaking the symmetry, which also means resonance is changed from BICs into (Q-BICs). **Figure 1c** shows the variation trend of each particle, the asymmetry parameter ($\alpha$) can be obtained by,

$$\alpha = \frac{a_L b_L - a_R b_R}{a_L b_L + a_R b_R} = \frac{S_L - S_R}{S_L + S_R}. \tag{1}$$

The length and width of left (right) nano brick are $a_L$ ($a_R$) and $b_L$ ($b_R$), respectively. The variation range of $a_L$ ($a_R$) is from 270 nm (90 nm) to 180 nm (180 nm); $b_L$ ($b_R$) is from 150 nm (50 nm) to 100 nm (100 nm). In the variation processes, the ratio of $a_L$:$b_L$ and $a_R$:$b_R$ are 9:5. The $D_x$ is fixed at 190 nm.

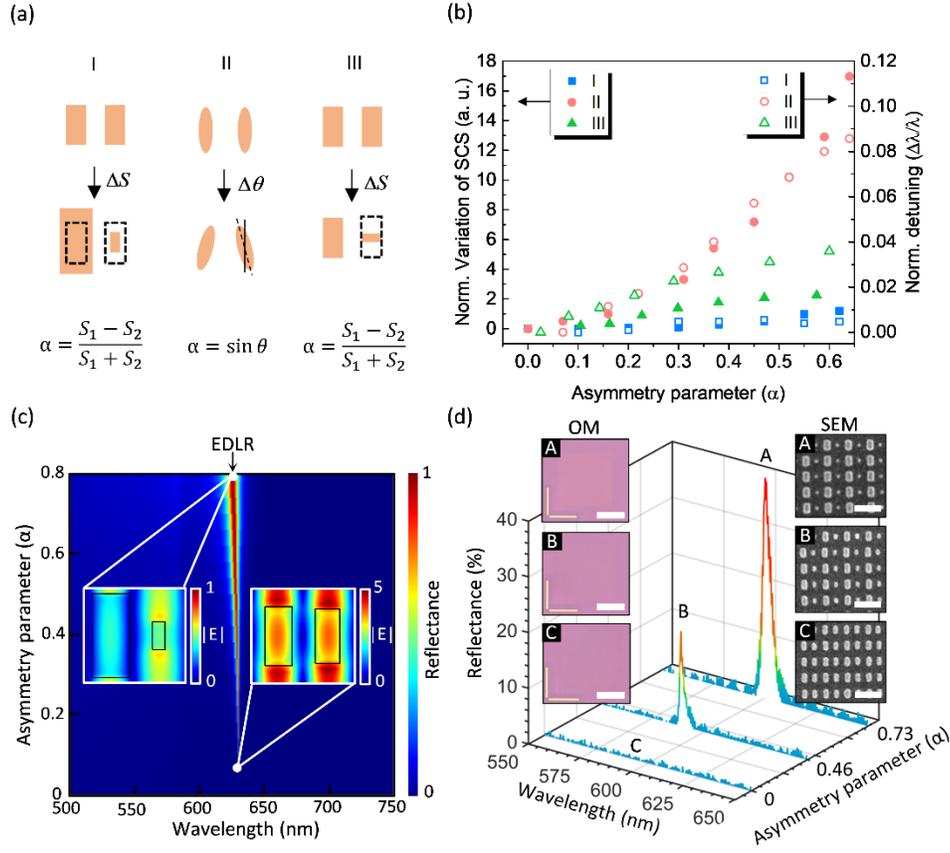

**Figure 2.** (a) Schematic of different designs of breaking symmetry. (b) Normalized scattering cross-section (solid marker) and shift of wavelength (hollow marker) versus different asymmetry parameters with different designs of breaking symmetry. (c) Simulated reflectance spectra in color mapping. The insets are the near-field distribution of $\alpha = 0.8$ and $\alpha = 0.05$. (d) Measured reflectance spectra of $\alpha = 0.73$ (structure A), $\alpha = 0.46$ (structure B), and $\alpha = 0$ (structure C). Insets show the optical reflection images and scanning electron microscopy images. The L shape marks in optical images are fabricated to locate these semi-transparent metasurface squares.

In order to compare the thresholds of laser, the wavelength of resonances have to be the same when asymmetry parameter is changed. To fix the resonance wavelength of different asymmetry parameter, left nano brick and right nano brick are changing at the same time to fix the scattering cross section and prevent the shift of resonance wavelength. **Figure 2a** shows different asymmetry designs, type I is the design which is used in our study, type II and III are general asymmetry design in other literatures[13,22,51]. Apparently, the resonance wavelength shift in type II and III is larger than type I, which is due to the large variation of scattering cross section of type II and II when changing the asymmetry parameters (**Figure 2b**).

The simulated reflectance spectra are plotted in **Figure 2c** by color mapping. In the simulation, the refractive index of surrounding medium is 1.46, and the period of each primitive unit cell in *x*- (*y*-) direction is 400 nm (300 nm). Obviously, the wavelengths of EDLR are fixed about 630 nm when the $\alpha$ is changed. Insets show the electric field distributions of $\alpha = 0.8$ and $\alpha = 0.05$. Compare the intensity of these electric field distributions, the intensity of $\alpha = 0.8$ is apparently weaker than that of $\alpha = 0.05$, which is because $\alpha = 0.05$ is close to BIC. The color bars of electric filed in **Figure 2c** are shown in different scale, which is because the intensity difference is too large. It is worth to note that the field distributions of $\alpha = 0$ is not shown in **Figure 2c**, which is because BICs cannot be excited by external excitation, which will be further discussed in the following paragraph.

**CHARACTERIZATION**

The measured reflectance spectra and scanning electron microscope (FEI Helios G3CX) images are shown in **Figure 2d** (please see the details in **Methods**). Structure A, B, C present the design of $\alpha = 0.73$, $\alpha = 0.46$, $\alpha = 0$. The spectra and images of reflectance were collected using an optical microscope (BX51, Olympus) and spectrometer (Kymera 193i, Andor) with a charge-coupled device (iVac 316 LDC-DD, Andor). To make the measurement close to normal incidence, the 5× objective lens (LMPLFLN 5X BD, numerical aperture [NA] = 0.13) was used to approximate 0-degree incidence. The reflectance spectra present sharp peaks when $\alpha$ is close to 0. However, the feature of BIC ($\alpha = 0$) cannot be observed in the reflectance spectrum. This phenomenon can also be observed by optical images in **Figure 2d**, which squares show the area of the metasurface. The metasurfaces gradually become transparent in reflection images when $\alpha$ is close to zero. The reason is that the external excitation is not able to excite BICs due to non-radiation channel in the BICs system.

In contrast, the internal excitation is able to excite the BICs, however, the feature still cannot be observed because of the same reason.

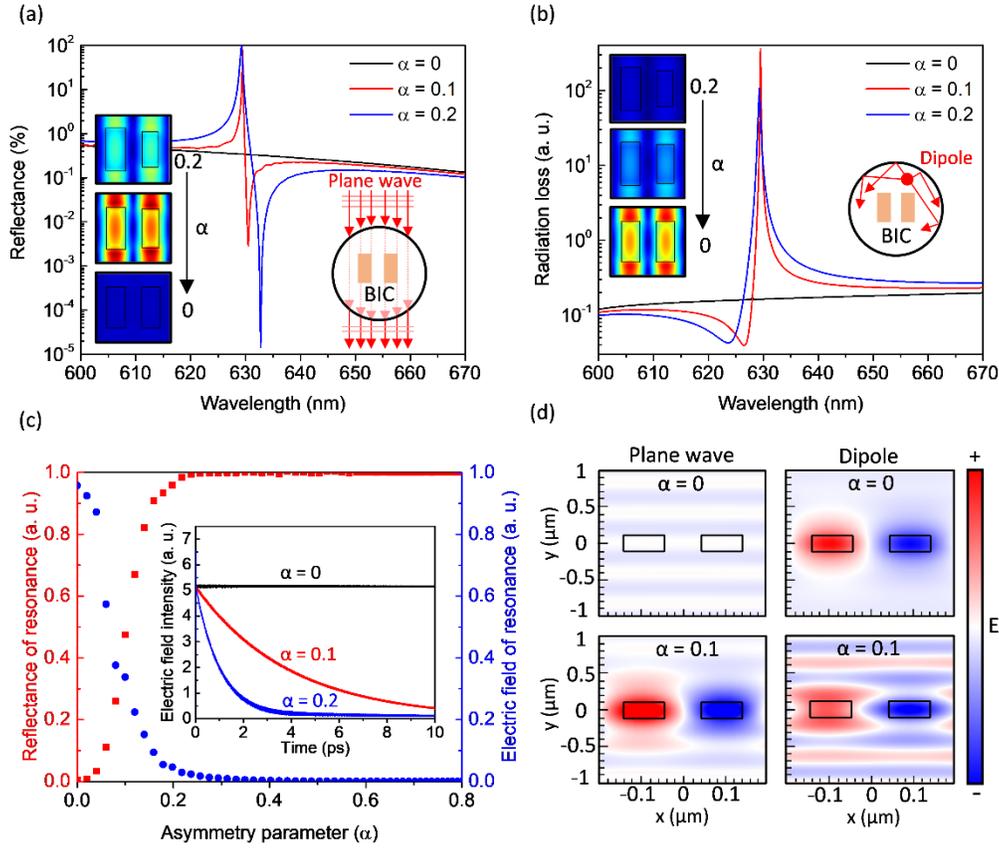

**Figure 3.** Simulated (a) external excitation spectra and (b) internal excitation spectra with different asymmetry parameters. Insets show the near-field distribution of different asymmetry parameters. (c) The plot of maximum reflectance and electric field intensity versus different asymmetry parameters. (d) Life-time of FDTD simulation of internal excitation with different asymmetry parameters.

**Figure 3a,b** shows the simulated spectra and electric field distributions of external excitation and that of internal excitation, respectively. In **Figure 3a**, the electric field strength is increased when $\alpha$ is decreased, however, the strength of electric field is completely vanished when $\alpha = 0$. To image the BIC is an ideal cavity, external energy cannot arrive inside of the ideal cavity, thus the near-field distribution of $\alpha = 0$ does not show

any localized field. In **Figure 3**, the reflectance spectra are shown in the log scale, and the Fano shape shows the constructive and destructive interferences. These two interferences are gradually close to each other when $\alpha$ close to zero; then these two interferences are overlapped, and show the flat spectrum without any feature. In contrast, **Figure 3b** shows the internal excitation by dipole sources. The electric field distributions show gradually increasing even $\alpha = 0$. However, the radiation loss spectra show the similar result with external excitation, which is because that an ideal cavity cannot absorb any energy, but neither emit any energy outside the system. In **Figure 3c**, the relation between radiation loss and localized field intensity can explain the mechanism from Q-BICs to BICs. This characteristic can also be observed by the decay time of modes in the inset. Comparing with BICs ($\alpha = 0$) and Q-BICs ($\alpha = 0.1, 0.2$), BICs show infinite decay time because of the property of lossless system. The simulated near-field distribution of BICs ($\alpha = 0$) and Q-BICs ($\alpha = 0.1$) with plane wave excitation and dipole excitation are shown in **Figure 3d**. By the plane wave excitation, BICs play a role of transparent object without any coupling, but in contrast, Q-BICs absorb the energy and excite the resonance successfully. By the dipole excitation, BICs traps all the energy without any radiation loss, and Q-BICs emit electro-magnetic wave to the free space.

To utilize the high Q-factor of BICs, PL measurement would be used to characterize the internal excitation source in the BICs system. Consider the resonance wavelength of our samples, R6G is a suitable candidate, which is because the wavelength of the peak in the spontaneous emission spectrum is relatively shorter than that of BIC (**Figure S1**). Therefore, the threshold of lasing is not affected by the re-absorption effect of R6G.

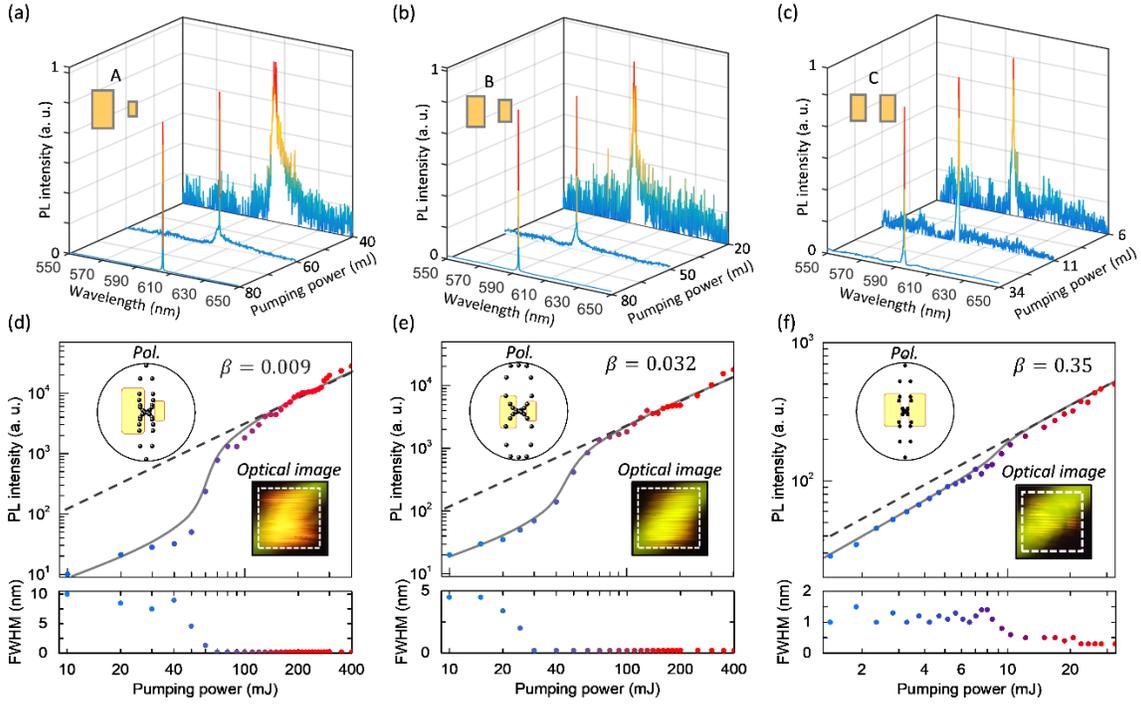

**Figure 4.** PL spectra of structure (a) A ($\alpha = 0.73$), (b) B ($\alpha = 0.46$), (c) C ($\alpha = 0$) with different pumping power which are close to lasing threshold. The L-L curve and linewidth evolution of structure (a) A ($\alpha = 0.73$), (b) B ($\alpha = 0.46$), (c) C ($\alpha = 0$). Insets show the optical images and polarization intensity, which are over the threshold.

To measure the PL and lasing spectra, the measurement setup is the same with reflectance measurement, but replace the pumping source to 532 nm pulse laser (MPL-III-532nm-1 $\mu$J-19101610, Changchun New Industries Optoelectronics Tech. Co.) (see the detail in **Method**). **Figure 4** shows the experimental results of PL signal. **Figures 4a,b,c**, which corresponds to structure A, B, C from **Figure 2**, respectively, show the PL spectra when the pumping power density is close to the lasing threshold. **Figures 4d,e,f** show the light in-light out curve (L-L curve) of structure A, B, C, respectively. Insets present the optical images and polarization intensity in which the pumping power is over the threshold, and the parallel fringes are the evidence that the mode of stimulated emission is assisted by EDLR. Interestingly, in the optical images, due to the very narrow resonance signal of Q-BICs, lasing signal can be observed in the almost transparent sample. In **Figure 4d,e,f** the lasing thresholds of Q-BICs is decreased when $\alpha$ is changed from 0.73 to 0, which is because the energy localized at gain medium is increased, then the $\beta$ is also increased. In principle, BICs are similar to perfect cavity because of the property of no-loss system. Based on the characteristic of no-loss system, the energy inside the BICs system cannot be radiated. Thus, the lasing signal of BIC should not be expected to be emitted. However, the lasing signal is still detected in the experiment, which is due to the defect in structures (**Figure S2**) and effects of finite arrays (**Figure S3**). The lasing area of structure A, B, and C are $100 \times 100$ $um^2$, $100 \times 100$ $um^2$, and $36 \times 25$ $um^2$, respectively. Apparently, the lasing area of structure C is much smaller than others, which is also because of the non-ideal of geometry. The oblique PL spectra are also measured to make sure the lasing mode is assisted by the EDLR (**Figure S4**).

**CONCLUSIONS**

In this article, the characteristics of BICs lasers are demonstrated by the simulation and the experiment in $Si_3N_4$ metasurfaces with R6G gain medium. $Si_3N_4$ metasurfaces with EDLR is utilized to generate Fano resonance and create BICs by complete dark hybrid-SLR. The complete dark SLR presents the infinite *Q*-factor and strong localized fields, which shows the property of BICs. Due to the BICs are lossless mode, the resonance cannot be observed and excited by external excitations. To excite the BICs, the gain material covered the $Si_3N_4$ metasurface, which can be treat as an internal excitation source. To apply the infinite *Q*-factor and strong localized field to laser, $Si_3N_4$ metasurfaces are covered by R6G. By combining with R6G, lasing can be observed in the normal direction (at $\Gamma$ point) successfully in the experiment, and this kind of setup could be further applied in microfluidics lasers. By tuning the *Q*-factor, the threshold of laser is decreased gradually when *Q*-factor is increased. Also, the low threshold laser with BIC is demonstrated by experiments with large beta. To analyze the lasing threshold of different structures, the design of breaking symmetry without the shift of resonance wavelength has been demonstrated. The results in this study provide a design rule for BIC metasurfaces with hybrid-SLR, and reveal the mechanism of BICs by internal and external excitations. For applications, results in this study can be further applied to high $Q$ lasers, optical sensing, non-linear optics enhancement, and single-photon sources.

**METHODS**

**Fabrication.** The plasma-enhanced chemical vapor deposition system (Oxford Plasmalab system 100) was utilized to deposit a $Si_3N_4$ thin film (thickness = 227 nm) on quartz substrate. Subsequently, a poly(methyl methacrylate) (PMMA)-A4 photoresist was spin-coated on the $Si_3N_4$ thin film for following lithography processes. Due to the exposure equipment is electron beam lithography (EBL) system, a conductive polymer (AR-PC 5090 by ALLRESIST) was spin-coated on the PMMA photoresist to avoid charge accumulation. After exposure by EBL, rectangular nanohole arrays are formed by development. To create etching mask, chromium layers (thickness = 20 nm) were deposited on a patterned PMMA layer, then lift-off by acetone to generate rectangular chromium array. To fabricate the $Si_3N_4$ metasurfaces, an inductively coupled plasma etching system (EIS-700 by ELIONIX) was used to etch the $Si_3N_4$ layer. Finally, the chromium layer was removed through wet etching.

**Measurement of PL spectra.** In this study, R6G PL material is used to play a role of gain material. The R6G liquid is deployed by R6G powder and Dimethyl sulfoxide (DMSO) liquid, and the concentration of R6G-DMSO liquid is 10 mM. To surround the $Si_3N_4$ metasurface by R6G-DMSO liquid, 4 $\mu$ liter R6G-DMSO is dropped on the $Si_3N_4$ metasurface, then a coverslip sandwiched R6G-DMSO between $Si_3N_4$ metasurface and coverslip. Utilize 532 nm pulse laser to pump the R6G and excite the PL signal of R6G. The repetition rate of pulse laser is about 4 kHz, pulse width is 1.24 ns, and the pumping spot size on the Si3N4 metasurface is about 150 um$^2$. The PL spectra were collected using an optical microscope (BX51, Olympus) and spectrometer (Kymera 193i, Andor) with a charge-coupled device (iVac 316 LDC-DD, Andor). The 5× objective lens (LMPLFLN 5X BD, numerical aperture [NA] = 0.13) was used to collect the signal from normal direction.

**ACKNOWLEDGEMENTS**

This work was supported by the Higher Education Sprout Project of the National Chiao Tung University and Ministry of Education and the Ministry of Science and Technology (MOST-107-2221-E-009-046-MY3; 107-2218-E-009-056; 108-2923-E-009-003-MY3). The research was also funded by Russian Foundation for Basic Research projects No. 19-52-52006.

**AUTHOR INFORMATIONS**

J.-H.Y. performed the sample fabrication, simulation and optical characterization. All authors joined the discussion and revised the manuscript.

**COMPETING INTERESTS STATEMENT**

The authors declare no competing financial interest.